\documentclass[12pt]{article}
\usepackage{amsmath}
\usepackage{amssymb}
\usepackage{array}
\usepackage{geometry}
\usepackage{graphicx}
\usepackage{enumerate}
\numberwithin{equation}{section}
\begin{document}
\vskip2cm\noindent
\begin{center}{\bf Mittag-Leffler Probability Density for Nonextensive Statistics and Superstatistics}\\
\vskip.3cm{\bf A.M. Mathai}\\
\vskip.1cm{Department of Mathematics and Statistics,}\\
\vskip0cm{McGill University, Montreal, Canada, H3A 2K6}\\
\vskip0cm{directorcms458@gmail.com}\\
\vskip.3cm {and}\\
\vskip.3cm{\bf H.J. Haubold}\\
\vskip0cm{Office for Outer Space Affairs, United Nations}\\
\vskip0cm{Vienna International Centre, A-1400 Vienna, Austria}\\
\vskip0cm{hans.haubold@gmail.com}\\
\end{center}

\vskip.5cm\noindent{\bf Abstract}\vskip.3cm It is shown that a Mittag-Leffler density has interesting properties. The Mittag-Leffler random variable has a structural representation in terms of a positive L\'evy variable and a power of a gamma variable where these two variables are independently distributed. It is shown that several central limit type properties hold but the limiting forms are positive L\'evy variable rather than a Gaussian variable. A path is constructed from a Mittag-Leffler function to the Mathai pathway model which also provides paths to nonextensive statistics and superstatistics.

\vskip.3cm\noindent{\bf Keywords:}\hskip.3cm Mittag-Leffler probability density, structural properties, central limit type results, attraction to positive L\'evy density, Mittag-Leffler function, Mathai pathway model, nonextensive statistics, superstatistics.

\vskip.5cm\noindent{\bf 1.\hskip.3cm Introduction } \vskip.3cm

The Mittag-Leffler function is a principal function in the area of fractional calculus, especially in fractional differential equations (Haubold, Mathai, and Saxena, 2011). A Mittag-Leffler density associated with a 3-parameter Mittag-Leffler function can be introduced in the following form (Mathai, 2005):

$$f(x)=\frac{x^{\alpha\beta-1}}{\delta^{\beta}}\sum_{k=0}^{\infty}\frac{(\beta)_k}{k!}\frac{(-1)^k(x^{\alpha})^k}{\delta^k\Gamma(\alpha\beta+\alpha k)},x\ge 0,\Re(\alpha)>0,\Re(\beta)>0,\delta>0.\eqno(1.1)
$$If $(\beta)_k$ is written as $(\beta)_k=\frac{\Gamma(\beta+k)}{\Gamma(\beta)}$ for $\Re(\beta)>0$ then we can rewrite the density as follows:

$$f(x)=\frac{x^{\alpha\beta-1}}{\delta^{\beta}\Gamma(\beta)}\sum_{k=0}^{\infty}\frac{\Gamma(\beta+k)}{k!}
\frac{(-1)^k(x^{\alpha})^k}{\delta^k\Gamma(\alpha\beta+\alpha k)},\Re(\beta)>0,\delta>0,\Re(\alpha)>0,x\ge 0.\eqno(1.2)
$$One can write (1.2) as a Mellin-Barnes integral as follows:
$$f(x)=\frac{1}{\delta^{\beta}\Gamma(\beta)}\frac{1}{2\pi i}\int_{c-i\infty}^{c+i\infty}\frac{\Gamma(s)\Gamma(\beta-s)}{\delta^{-s}\Gamma(\alpha\beta-\alpha s)}x^{\alpha\beta-1-\alpha s}{\rm d}s,0<\Re(s)<\Re(\beta).\eqno(1.3)
$$Let us rewrite this Mellin-Barnes integral in a more convenient form. Put $\alpha\beta-1-\alpha s=-s_1$. Then

$$f(x)=\frac{1}{\Gamma(\beta)}\frac{1}{2\pi i}\int_{c_1-i\infty}^{c_1+i\infty}\frac{\Gamma(\beta-\frac{1}{\alpha}+\frac{s_1}{\alpha})\Gamma(1+\frac{1}{\alpha}
-\frac{s_1}{\alpha})}
{\delta^{\frac{1}{\alpha}-\frac{s_1}{\alpha}}\Gamma(2-s_1)}x^{-s_1}{\rm d}s_1\eqno(1.4)
$$for $-\Re(\alpha\beta)+1<c_1<\Re(\alpha)+1$, $-\Re(\alpha\beta)+1<\Re(s)<\Re(\alpha)+1<2,0<\Re(\alpha)\le 1$. If $f(x)$ is a density then the integrand in (1.4) must be this density's Mellin transform or written in terms of expected values
$$E(x^{s-1})=\frac{1}{\Gamma(\beta)}\frac{\Gamma(\beta-\frac{1}{\alpha}+\frac{s}{\alpha})\Gamma(1+\frac{1}{\alpha}-
\frac{s}{\alpha})}{\delta^{\frac{1}{\alpha}-\frac{s}{\alpha}}\Gamma(2-s)}\eqno(1.5)
$$for $0<\Re(\alpha)\le 1,-\Re(\alpha\beta)+1<\Re(s)<\Re(\alpha)+1<2$. Note that when $s\to 1$ the right side in (1.5) is $1$. This observation, together with the observation that $f(x)\ge 0$, is also a test for showing that $f(x)$ in (1.1) is a density.

\vskip.3cm\noindent{\bf 2.\hskip.3cm Properties of Mittag-Leffler Probability Density}

\vskip.3cm
Let us compute the Laplace transform of the density $f(x)$, with Laplace parameter $s$.
\begin{align*}
E({\rm e}^{-sx})&=\int_0^{\infty}{\rm e}^{-sx}f(x){\rm d}x\\
&=\frac{1}{\delta^{\beta}}\sum_{k=0}^{\infty}\frac{(\beta)_k}{k!}\frac{(-1)^k}{\delta^k}
\int_0^{\infty}\frac{x^{\alpha\beta+\alpha k-1}{\rm e}^{-sx}}{\Gamma(\alpha\beta+\alpha k)}{\rm d}x\\
&=\frac{1}{\delta^{\beta}}\sum_{k=0}^{\infty}\frac{(\beta)_k}{k!}\frac{(-1)^k}{\delta^k}s^{-(\alpha\beta+\alpha k)}=(1+\delta s^{\alpha})^{-\beta}&(2.1)\end{align*}
for $|\delta s^{\alpha}|<1$. This (2.1) looks like the Laplace transform of a gamma density with $s$ replaced by $s^{\alpha}$. This gives a decomposition of a Mittag-Leffler variable. Consider a real random variable $y$ having the structure $y=uv^{\frac{1}{\alpha}}$ where $u$ and $v$ are independently distributed with $u$ having a positive L\'evy density with parameter $\alpha$ and $v$ having a gamma density with shape parameter $\beta$ and scale parameter $\delta$ or with the density

$$g(v)=\frac{v^{\beta-1}{\rm e}^{-\frac{v}{\delta}}}{\delta^{\beta}\Gamma(\beta)},v\ge 0,\delta>0,\beta>0\eqno(2.2)
$$and zero elsewhere. Then
\begin{align*}
E(v^{\frac{1}{\alpha}})^{s-1}&=\frac{1}{\delta^{\beta}\Gamma(\beta)}\int_0^{\infty}
v^{\frac{s}{\alpha}-\frac{1}{\alpha}+\beta-1}{\rm e}^{-\frac{v}{\delta}}{\rm d}v\\
&=\frac{\delta^{\beta+\frac{s}{\alpha}-\frac{1}{\alpha}}\Gamma(\beta+\frac{1}{\alpha}+\frac{s}{\alpha})}{\delta^{\beta}
\Gamma(\beta)}=\frac{\delta^{\frac{s}{\alpha}-\frac{1}{\alpha}}
\Gamma(\beta+\frac{s}{\alpha}-\frac{1}{\alpha})}{\Gamma(\beta)}&(2.3)\end{align*}for $\delta>0,\Re(\beta)>0,\Re(s)>1-\Re(\alpha\beta)$. For a positive L\'evy density the Laplace transform, with Laplace parameter $s$, is given by ${\rm e}^{-s^{\alpha}}$. Let us consider the Laplace transform of $y$ with Laplace parameter $s$. That is, $E({\rm e}^{-sy})$. This can be written in terms of a conditional expectation. That is,

$$E({\rm e}^{-sy})=E[E({\rm e}^{-suv^{\frac{1}{\alpha}}}|_{v})].\eqno(2.4)
$$But the conditional expectation of $y$ , for given $v$, is the Laplace transform of a positive L\'evy density with Laplace parameter $sv^{\frac{1}{\alpha}}$. Then this conditional expectation is given by the following:
$$E[{\rm e}^{-sy}|_{v}]=E[{\rm e}^{-suv^{\frac{1}{\alpha}}}|_{v}]={\rm e}^{-(sv^{\frac{1}{\alpha}})^{\alpha}}={\rm e}^{-s^{\alpha}v}.\eqno(2.5)
$$Now, taking the outside expectation with respect to $v$, we have

$$E[{\rm e}^{-sy}]=E[{\rm e}^{-v(s^{\alpha})}].
$$But this is the Laplace transform of a gamma density with Laplace parameter $s^{\alpha}$. Therefore,
$$E[{\rm e}^{-sy}]=[1+\delta s^{\alpha}]^{-\beta},|\delta s^{\alpha}|<1.\eqno(2.6)
$$This means that a Mittag-Leffler random variable $x$ has the structural representation
$$x=uv^{\frac{1}{\alpha}}\eqno(2.7)
$$where $u$ and $v$ are independently distributed with $u$ having a positive L\'evy distribution and $v$ having a gamma distribution with shape parameter $\beta$ and scale parameter $\delta$. Due to statistical independence
$$E(x^{s-1})=E(u^{s-1})E(v^{\frac{1}{\alpha}})^{s-1}.\eqno(2.8)
$$But from (2.3)

$$E[(v^{\frac{1}{\alpha}})^{s-1}]=\frac{\delta^{\frac{s}{\alpha}-\frac{1}{\alpha}}\Gamma(\beta+\frac{s}{\alpha}
-\frac{1}{\alpha})}{\Gamma(\beta)},
$$for $\delta>0, \Re(\beta)>0, \Re(s)>1-\Re(\alpha\beta)$. Hence, from (1.5)

$$E[u^{s-1}]=\frac{\Gamma(1+\frac{1}{\alpha}-\frac{s}{\alpha})}{\Gamma(2-s)},\eqno(2.9)
$$for $0<\Re(\alpha)\le 1,\Re(s)<1+\Re(\alpha)<2.$ Thus, (2.9) gives the Mellin transform of a positive L\'evy density with parameter $\alpha$. We have derived the Laplace transform of the Mittag-Leffler density in (1.1) as
$$L_f(s)=(1+\delta s^{\alpha})^{-\beta}, |\delta s^{\alpha}|<1.\eqno(2.10)
$$Hence, if $\delta$ is replaced by $\frac{\delta}{\beta}$ and take the limit when $|\beta|\to\infty$ then we have
$$\lim_{|\beta|\to\infty}[1+\frac{\delta}{\beta}s^{\alpha}]^{-\beta}={\rm e}^{-\delta s^{\alpha}},\mbox{ for fixed }\delta,\alpha\eqno(2.11)
$$which shows that when $\delta$ is replaced by $\frac{\delta}{\beta}$ then a Mittag-Leffler density goes to a L\'evy density or the Mittag-Leffler random variables goes to $\delta^{\frac{1}{\alpha}}$ times a L\'evy variable when $|\beta|\to\infty$.

\vskip.3cm\noindent{\bf 3.\hskip.3cm Central Limit Type Results}

\vskip.3cm Let $x_1,x_2,...,x_n$ be a simple random sample from the Mittag-Leffler population (1.1) or let $x_1,...,x_n$ be independently and identically distributed as (1.1). Then $u=x_1+...+x_n$ has the Laplace transform, denoted by $L_u(s)$, as
$$L_u(s)=[1+\delta s^{\alpha}]^{-n\beta}.
$$Then $w=\frac{u}{n^{\frac{1}{\alpha}}}=\frac{x_1+...+x_n}{n^{\frac{1}{\alpha}}}$ has the Laplace transform
$$L_w(s)=[1+\frac{\delta}{n}s^{\alpha}]^{-n\beta}\to {\rm e}^{-\delta\beta s^{\alpha}}\mbox{ as }n\to\infty, \mbox{ for fixed }\delta,\beta .\eqno(2.12)
$$If $\delta$is replaced by $\frac{\delta}{\beta}$ then
$$L_{w}(s)\to {\rm e}^{-s^{\alpha}}
$$or we get the positive L\'evy variable, as $n\to\infty$. 
\vskip.2cm Let $u_1,...,u_n$ be independently and identically distributed positive L\'evy variable with parameter $\alpha$. Then $t=\frac{u_1+...+u_n}{n^{\frac{1}{\alpha}}}$ is again a positive L\'evy variable with parameter $\alpha$ for every given $n$. Then $x=tv^{\frac{1}{\alpha}}$, for every given $n$. This is another structural representation of a Mittag-Leffler random variable. Let $u_1,...,u_n$ be independently and identically distributed positive L\'evy variables with parameter $\alpha$ and let $v_1,...,v_n$ be independently and identically distributed gamma random variables with shape parameter $\beta$ and scale parameter $\delta$, and let the $u_j$'s and $v_j$'s be independently distributed, then $\frac{1}{n^{\frac{1}{\alpha}}}[u_1v_1^{\frac{1}{\alpha}}+...+u_nv_n^{\frac{1}{\alpha}}]$ has a positive L\'evy density with parameter $\alpha$ when $n\to\infty$.

\vskip.3cm\noindent{\bf 3.1.\hskip.3cm Series representation of a L\'evy density}

\vskip.3cm From (2.9)
$$E(u^{s-1})=\frac{\Gamma(1+\frac{1}{\alpha}-\frac{s}{\alpha})}{\Gamma(2-s)},\Re(s)<1+\Re(\alpha)<2
.$$Let $g_1(u)$ be the density of $u$. Then
$$g_1(u)=\frac{1}{2\pi i}\int_{c-i\infty}^{c+i\infty}\frac{\Gamma(1+\frac{1}{\alpha}-\frac{s}{\alpha})}{\Gamma(2-s)}u^{-s}{\rm d}s,0<c<1+\Re(\alpha).
$$The poles of $\Gamma(1+\frac{1}{\alpha}-\frac{s}{\alpha})$ are at $1+\frac{1}{\alpha}-\frac{s}{\alpha}=-\nu,\nu=0,1,2,...$ The residue at this pole is given by

$$\lim_{s\to 1+\alpha(1+\nu)}[1+\frac{1}{\alpha}-\frac{s}{\alpha}+\nu)\frac{\Gamma(1+\frac{1}{\alpha}-\frac{s}{\alpha})u^{-s}}{\Gamma(2-s)}
=\frac{(-1)^{\nu}}{\nu!}\frac{u^{-(1+\alpha(1+\nu))}}{\Gamma(1-\alpha-\alpha\nu)}
$$for $0<\Re(\alpha)<1,u>1$. Therefore
$$g_1(u)=\frac{1}{u^{1+\alpha}}\sum_{\nu=0}^{\infty}\frac{(-1)^{\nu}}{\nu!}\frac{u^{-\alpha\nu}}
{\Gamma(1-\alpha-\alpha\nu)},u>1,0<\Re(\alpha)<1
$$for $1-\alpha-\alpha\nu\ne 0,-1,...$.

\vskip.3cm\noindent{\bf 4.\hskip.3cm From Mittag-Leffler Function to Mathai Pathway Model, Nonextensive Statistics and Superstatistics}

\vskip.3cm Consider the following function associated with a 3-parameter Mittag-Leffler function.

$$f^{*}(x)=c~x^{\eta}\sum_{k=0}^{\infty}\frac{(\gamma)_k}{k!}\frac{(x^{\alpha})^k(-\delta)^k}{\Gamma(\beta+\alpha k)}.$$for $\Re(\alpha)>0,\Re(\beta)>0$. Consider $\Gamma(\beta)f^{*}(x\beta)$. Let $|\beta|\to\infty$. Then
$$\lim_{|\beta|\to\infty}\frac{\Gamma(\beta)f^{*}(x\beta)}{\beta^{\eta}}=cx^{\eta}\sum_{k=0}^{\infty}\frac{(\gamma)_k}{k!}x^{\alpha k}(-\delta)^k\lim_{|\beta|\to\infty}\frac{\Gamma(\beta)\beta^{\alpha k}}{\Gamma(\beta+\alpha k)}.
$$By using Stirling's approximation we have
$$\frac{\Gamma(\beta)
\beta^{\alpha k}}{\Gamma(\beta+\alpha k)}
\approx \frac{\sqrt{2\pi}\beta^{\beta+\alpha k-\frac{1}{2}}{\rm e}^{-\beta}}{\sqrt{2\pi}\beta^{\beta+\alpha k-\frac{1}{2}}{\rm e}^{-\beta}}=1.$$Hence when $|\beta|\to\infty$
$$\Gamma(\beta)f^{*}(\beta x)\to c~x^{\eta}\sum_{k=0}^{\infty}\frac{(\gamma)_k}{k!}(-\delta x^{\alpha})^k=c~x^{\eta}[1+\delta x^{\alpha}]^{-\gamma},\mbox{ for fixed }\delta,\gamma.
$$This, in fact, is the pathway model. Take $\delta=a(q-1),q>1$ and $\gamma=\frac{1}{q-1},a>0$ then $\Gamma(\beta)f^{*}(\beta x)$ goes to
$$f_2(x)=c_2x^{\eta}[1+a(q-1)x^{\alpha}]^{-\frac{1}{q-1}},q>1,a>0.\eqno(4.1)
$$This is the pathway model for $q>1$. Writing $q-1=-(1-q),q<1$ and then taking the limit when $q\to 1$ we get the other two forms of the pathway model, namely
$$f_1(x)=c_1x^{\eta}[1-a(1-q)x^{\alpha}]^{\frac{1}{1-q}},q<1,a>0\eqno(4.2)
$$and
$$f_3(x)=c_3 x^{\eta}{\rm e}^{-ax^{\alpha}},a>0\eqno(4.3)
$$where $c_1,c_2,c_3$ are constants. If $f_1,f_2,f_3$ are to be treated as statistical densities then $c_1,c_2,c_3$ are the normalizing constants there.  Here (4.1) for $a=1,q>1$ and $q\to 1$, but not for $q<1$, is superstatistics in statistical mechanics,  Beck (2004)  and Cohen (2004). Taking $\eta=0,a=1,\alpha=1$ in (4.1) we get nonextensive statistics in statistical mechanics, Tsallis (2009), for all cases $q<1,q>1,q\to 1$. Nonextensive statistics is also a power function model in the sense
$$\frac{{\rm d}}{{\rm d}x}f_1(x)=-[f_1(x)]^{q}$$
where $f_1(x)$ is that of (4.2) for $\eta=0,a=1,\alpha=1$ and $q<1,q>1,q\to 1$.
\vskip.2cm
The advantage of the pathway model introduced in Mathai (2005), a special case of which for the real scalar positive variable is the case in (4.2), is that it can switch around to three functional forms. For $q<1$ it stays in the generalized type-1 beta family of functions, for $q>1$ it switches into generalized type-2 beta family of functions and for $q\to 1$ both the above forms go to the generalized gamma family of functions. Hence if the generalized gamma family is the ideal situation or the stable situation in a physical system then the pathway model can capture the unstable or chaotic neighborhood as well as the the path leading to the stable situation. For other pathways see, for example, Mathai and Haubold (2007, 2011, 2018).

\vskip.3cm\noindent
\begin{center}{\bf References}
\end{center}
\vskip.2cm\noindent C. Beck, Superstatistics: Theory and applications, {\it Continuum Mechanics and Thermodynamics}, {\bf 16}(2004), 293-304.
\vskip.2cm\noindent E.G.D. Cohen, Superstatistics, {\it Physica D: Nonlinear Phenomena}, {\bf 193}(2004), 35-52.
\vskip.2cm\noindent H.J. Haubold, A.M. Mathai, and R.K. Saxena, Mittag-Leffler functions and their applications, {\it Journal of Applied Mathematics}, 2011, 298628, 51 pages..
\vskip.2cm\noindent A.M. Mathai, A pathway to matrix-variate gamma and normal densities, {\it Linear Algebra and its Applications}, {\bf 396}(2005), 317-328.
\vskip.2cm\noindent A.M. Mathai and H.J. Haubold, On generalized entropy measure and pathways, {\it Physica A}, {\bf 385}(2007), 493-500.
\vskip.2cm\noindent A.M. Mathai and H.J. Haubold, A pathway from Bayesian statistical analysis to superstatistics, {\it Applied Mathematics and Computations}, {\bf 218}(2011), 799-804.
\vskip.2cm\noindent C. Tsallis, {\it Introduction to Nonextensive Statistical Mechanics: Approaching a Complex World}, Springer, New York, 2009.
\vskip.2cm\noindent A.M. Mathai and H.J. Haubold, {\it Erd\'elyi-Kober Fractional Calculus: From a Statistical Perspective, Inspired by Solar Neutrino Physics}, Springer Briefs in Mathematical Physics {\bf 31}, Springer Nature Singapore, 2018.

\end{document}